\documentclass[Physsubmission, Phys]{SciPost}
\hypersetup{
	colorlinks,
	linkcolor={red!50!black},
	citecolor={blue!50!black},
	urlcolor={blue!80!black}
}
\usepackage{braket}
\usepackage[bitstream-charter]{mathdesign}
\DeclareSymbolFont{usualmathcal}{OMS}{cmsy}{m}{n}
\DeclareSymbolFontAlphabet{\mathcal}{usualmathcal}

\begin{document}

\begin{center}{\Large \textbf{
T-odd quark-gluon-quark correlation function
in the light-front quark-diquark model\\
}}\end{center}

% TODO: write the author list here. Use initials + surname format.
% Separate subsequent authors by a comma, omit comma at the end of the list.
% Mark the corresponding author with a superscript *.
\begin{center}
Shubham Sharma\textsuperscript{1},
Narinder Kumar\textsuperscript{2} and
Harleen Dahiya\textsuperscript{1$\star$}
\end{center}

\begin{center}
{\bf 1} Department of Physics, Dr. B. R. Ambedkar National Institute of Technology, Jalandhar 144011, India
\\
{\bf 2} Department of Physics, Doaba College, Jalandhar 144004, India
\\

* dahiyah@nitj.ac.in

\end{center}

\begin{center}
\today
\end{center}

% For convenience during refereeing (optional),
% you can turn on line numbers by uncommenting the next line:
%\linenumbers

\definecolor{palegray}{gray}{0.95}
\begin{center}
\colorbox{palegray}{
  \begin{tabular}{rr}
  \begin{minipage}{0.1\textwidth}
    \includegraphics[width=22mm]{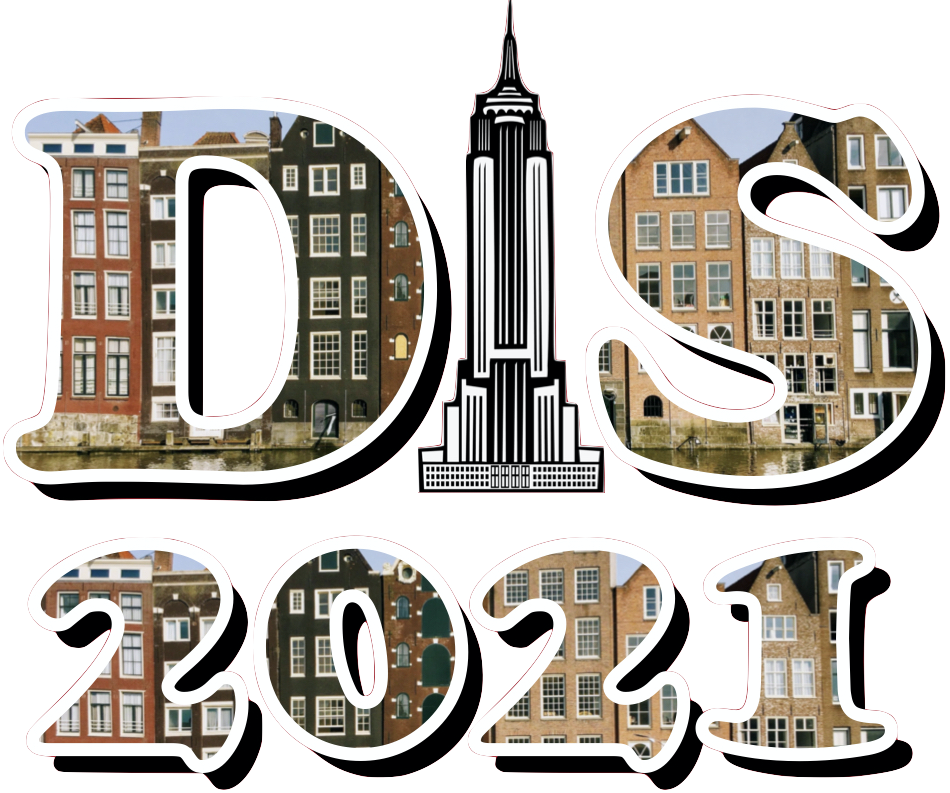}
  \end{minipage}
  &
  \begin{minipage}{0.75\textwidth}
    \begin{center}
    {\it Proceedings for the XXVIII International Workshop\\ on Deep-Inelastic Scattering and
Related Subjects,}\\
    {\it Stony Brook University, New York, USA, 12-16 April 2021} \\
    \doi{10.21468/SciPostPhysProc.?}\\
    \end{center}
  \end{minipage}
\end{tabular}}
\end{center}
%_________________________Lu2012____(edited)____________________________%
\section*{Abstract}
{\bf We have scrutinized the transverse momentum dependent quark–gluon–quark correlation function. We have utilized the light-front quark-diquark model to study the time-reversal-odd interaction-dependent twist-3 gluon distributions which we have obtained from the disintegration of the transverse momentum dependent quark–gluon–quark correlator. Specifically, we have studied the behavior of $\tilde{e}_{L}$ and $\tilde{e}_{T}$ while considering our diquark to be an axial-vector.}
%____________________________Abstract Done______________________________%

\section{Introduction}
\label{sec:intro}

%_________________________Lu2012___________(edited)______________________%
Theoretically, in the description of hadron containing (semi-)inclusive high energy processes, the cross-sections are generally written in the powers of $1 / Q$ with $Q$ being the large momentum transfer of the collision. The convolution of the hard scattering coefficients and leading-twist distribution functions is used to communicate the contribution at the leading power. In the $1 / Q$ expansion,  the first sub leading power of twist-3 distribution and/or fragmentation functions are added to the cross-section \cite{luref1}. 
%_________________________________________________________%
%_________________________Lu2012____(edited)____________________________%
Twist-3 distribution functions are not illuminated in the form of probability which is obvious and contrary to the interpretation of the twist-2 distributions. They are rather used to describe the parton densities in the core the nucleon. Information about the nucleon's parton structure is obtained from twist-3 distribution functions \cite{luref2}, particularly when the parton transverse momenta exist. The appeal on the twist-3 contributions also comes from the fact that they are related to the multi-parton correlation in the interior of nucleon \cite{luref3, luref4}.

%_______________________________Lu2012__________________________%

%_____________________(edited)_______________________________%

In this paper, we have utilized quark-diquark model to examine the quark distributions of twist-3 level which are ciphered in the quark-gluon-quark correlation. We have emphasized on the time-reversal-odd (T-odd) transverse momentum dependent  distributions (TMDs). In the single-spin asymmetries (SSAs) measured in semi-inclusive deeply inelastic scattering (SIDIS) \cite{{luref9,luref10,luref11}} the leading twist T-odd TMDs \cite{luref5} play important roles in the TMD factorization approach \cite{luref6,luref7,luref8}. In the TMD factorization approach at the twist-3 level there are eight T-odd distributions that can contribute to various azimuthal asymmetries in the SIDIS \cite{luref2} and Drell–Yan \cite{luref12} processes. Even though the  twist-3 contributions are suppressed due to $1 / Q$, still these experimental observables have potential and may be accessible in the kinematical regime where $Q$ is not so large. The ideal experiments for exploring this kinematical region are at  PAX \cite{luref15} and Jefferson Lab \cite{luref13,luref14}.
%_________________________________________________________________________%

\section{Light-Front Quark-Diquark Model}

%_____________________raw maji 2017______model1 pic______________%

%_______________________________edited___________________________________%

We contemplate on our problem by considering the light-front quark-diquark model  \cite{majiref21}, where the proton has a spin-flavor $S U(4)$ structure and is written as a combination of isoscalar-scalar diquark singlet $\left|u S^{0}\right\rangle$, isoscalarvector diquark $\left|u A^{0}\right\rangle$ and isovector-vector diquark $\left|d A^{1}\right\rangle$ states \cite{luref22 , luref32 }. We have 
\begin{equation}
|P ; \pm\rangle=C_{S}\left|u S^{0}\right\rangle^{\pm}+C_{V}\left|u A^{0}\right\rangle^{\pm}+C_{V V}\left|d A^{1}\right\rangle^{\pm},
\end{equation}
where $S$ and $A$ denote the scalar and axial-vector diquark and their superscripts are used to depict the isospin of that diquark. Here, we have used the light-cone convention $x^{\pm}=x^{0} \pm x^{3}$ and the frame is chosen such that the proton has no transverse momentum, i.e., $P \equiv\left(P^{+}, \frac{M^{2}}{P^{+}}, \mathbf{0}_{\perp}\right)$; where the struck quark and diquark have momentum $p \equiv$ $\left(x P^{+}, \frac{p^{2}+\left|\mathbf{p}_{\perp}\right|^{2}}{x P^{+}}, \mathbf{p}_{\perp}\right) \quad$ and $\quad P_{X} \equiv$ $((1-x) P^{+}, P_{X}^{-},-\mathbf{p}_{\perp})$,
respectively. $x=p^{+} / P^{+}$ is used to denote the longitudinal momentum fraction carried by the struck quark.
The two particle Fock-state expansion for axial-vector diquark is given as \cite{majiref25}
\begin{equation}
\begin{aligned}|\nu A\rangle^{\pm}=& \int \frac{d x d^{2} \mathbf{p}_{\perp}}{2(2 \pi)^{3} \sqrt{x(1-x)}}
%\\ &%
\times\left[\psi_{++}^{r(\nu)}\left(x, \mathbf{p}_{\perp}\right)\left|+\frac{1}{2}+1 ; x P^{+}, \mathbf{p}_{\perp}\right\rangle\right.
\\ &+\psi_{-+}^{r(\nu)}\left(x, \mathbf{p}_{\perp}\right)\left|-\frac{1}{2}+1 ; x P^{+}, \mathbf{p}_{\perp}\right\rangle 
%\\ &%
+\psi_{+0}^{r(\nu)}\left(x, \mathbf{p}_{\perp}\right)\left|+\frac{1}{2} 0 ; x P^{+}, \mathbf{p}_{\perp}\right\rangle \\ &+\psi_{-0}^{r(\nu)}\left(x, \mathbf{p}_{\perp}\right)\left|-\frac{1}{2} 0 ; x P^{+}, \mathbf{p}_{\perp}\right\rangle 
%\\ &%
+\psi_{+-}^{r(\nu)}\left(x, \mathbf{p}_{\perp}\right)\left|+\frac{1}{2}-1 ; x P^{+}, \mathbf{p}_{\perp}\right\rangle \\ & \left.+\psi_{-}^{r(\nu)}\left(x, \mathbf{p}_{\perp}\right)\left|-\frac{1}{2}-1 ; x P^{+}, \mathbf{p}_{\perp}\right\rangle\right], \end{aligned}
\end{equation}
\\
where $\left|\lambda_{q} \lambda_{D} ; x P^{+}, \mathbf{p}_{\perp}\right\rangle$ represents the two-particle state with quark helicity $\lambda_{q}=\pm \frac{1}{2}$, the helicity vector diquark is  $\lambda_{D}=\pm 1,0$ (triplet). The superscript of $\psi$, $r$ denotes the helicity of nucleon and the flavor index for the flavors $u$ and $d$ is denoted by $\nu$.

\section{Transverse Momentum Dependent Distributions (TMDs)}
\label{sec:another}
%There is no strict length limitation, but the authors are strongly encouraged to keep contents to the strict minimum necessary for peers to reproduce the research described in the paper.

\subsection{Quark-gluon-quark correlation function}
%You are free to use dividers as you see fit.

%____________________________Cal 1a___Lu2012____edited_________________%
We start our calculations with the transverse momentum dependent quark-gluon-quark correlation function which is defined in \cite{luref16}

\begin{equation}
\begin{aligned}
\left(\tilde{\Phi}_{A}^{[\pm] \alpha}\right)_{i j}\left(x, p_{T}\right) \equiv \int \frac{d^{2} \xi_{T} d \xi^{-}}{(2 \pi)^{3}} e^{i p \xi}
\times\langle P, S| \bar{\psi}_{j}(0) g \int_{\pm \infty}^{\xi^{-}} d \eta^{-} \mathcal{L}^{[\pm]}\left(0, \eta^{-}\right) \\
F^{+\alpha}(\eta) \left.\mathcal{L}^{\xi_{T}, \xi^{+}}\left(\eta^{-}, \xi^{-}\right) \psi_{i}(\xi)|P, S\rangle_{c}\right|_{\eta^{+}=\xi^{+}=0,\eta_{T}=\xi_{T},p^{+}=x P^{+}},
\end{aligned}
\end{equation}
where $F^{\mu v}$ is the  gluon antisymmetric field strength tensor. By definition, gauge-invariance is certified by the gauge-links $\mathcal{L}^{[\pm]}$ and $\mathcal{L}^{\xi_{T}, \xi^{+}}$. The sign " $\pm$ " in the subscript or superscript represents the future/past-pointing \cite{luref7} nature gauge-link between the quark and gluon in the SIDIS/Drell-Yan processes, respectively.
%_________________________________________________________________________%

%_____________________________Cal 2__Lu2012_________edited______________%
The correlator can be rewritten further as \cite{lu2012}

\begin{equation}
\begin{aligned}
\left(\tilde{\Phi}_{A}^{[\pm] \alpha}\right)_{i j}\left(x, p_{T}\right)=i g \int \frac{d^{2} \xi_{T} d \xi^{-} d \eta^{-}}{(2 \pi)^{4}} \int d x^{\prime} \frac{e^{i x^{\prime} p+\eta^{-}}}{\left(x^{\prime} \mp i \epsilon\right)}
\quad \times e^{i\left[\left(x-x^{\prime}\right) P^{+} \cdot \xi^{-}-p_{T} \cdot \xi_{T}\right]}\\
\langle P, S| \bar{\psi}_{j}(0) \mathcal{L}^{[\pm]}\left(0, \eta^{-}\right) F^{+\alpha}(\eta)
\quad \times\left.\mathcal{L}^{\xi_{T}, \xi^{+}}\left(\eta^{-}, \xi^{-}\right) \psi_{i}(\xi)|P, S\rangle\right|_{\eta^{+}=\xi^{+}=0, \eta_{T}=\xi_{T}}, 
\end{aligned}
\end{equation}

where the factor $1 /\left(x^{\prime} \mp i \epsilon\right)$ in Eq. (4) can be written as
\begin{equation}
\frac{1}{\left(x^{\prime} \mp i \epsilon\right)}=\mathrm{P}\left(\frac{1}{x^{\prime}}\right) \pm i \delta\left(x^{\prime}\right).
\end{equation}
%________________________Figure_______________________________________________%

\begin{figure}[h]
\centering
\includegraphics[width=0.67\textwidth]{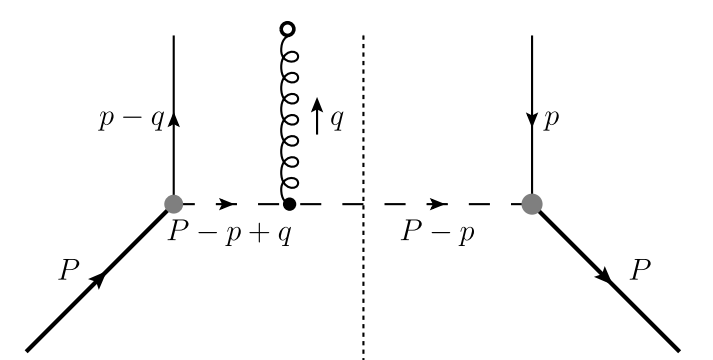}
\caption{Diagram corresponding to quark-gluon-quark correlator. $P$, $p$ and $q$ denote the nucleon, quark and diquark momenta respectively \cite{lu2012, luref20}.}
\label{ref}
\end{figure}

%_______________________________________________________________________%
\subsection{T-Odd Twist-3 Distributions}

We can break the quark-gluon-quark correlator as \cite{luref2}
\begin{equation}
\begin{aligned}
\tilde{\Phi}_{A}^{\alpha}\left(x, p_{T}\right)=& \frac{x M}{2}\left\{\left[\left(\tilde{f}^{\perp}-i \tilde{g}^{\perp}\right) \frac{p_{T \rho}}{M}-\left(\tilde{f}_{T}^{\prime}+i \tilde{g}_{T}^{\prime}\right) \epsilon_{T \rho \sigma} S_{T}^{\sigma}\right.\right.
\left.-\left(\tilde{f}_{s}^{\perp}+i \tilde{g}_{s}^{\perp}\right) \frac{\epsilon_{T \rho \sigma} p_{T}^{\sigma}}{M}\right]\\
&\left(g_{T}^{\alpha \rho}-i \epsilon_{T}^{\alpha \rho} \gamma_{5}\right)-\left(\tilde{h}_{s}+i \tilde{e}_{s}\right) \gamma_{T}^{\alpha} \gamma_{5}+[(\tilde{h}+i \tilde{e})
\left.\left.+\left(\tilde{h}_{T}^{\perp}-i \tilde{e}_{T}^{\perp}\right) \frac{\epsilon_{T}^{\rho \sigma} p_{T \rho} S_{T \sigma}}{M}\right] i \gamma_{T}^{\alpha}\right\} \frac{\not x_{+}}{2} .
\end{aligned}
\end{equation}
This equation contains interaction-dependent twist-3 quark distributions, which is in turn dependent on the longitudinal momentum fraction and the transverse momentum denoted by $x$ and $p_{T}$ respectively. These are denoted by the functions appearing with a tilde.
Out of these, 
$\tilde{g}^{\perp}$, $\tilde{f}_{T}$ (or $\left.\tilde{f}_{T}^{\prime}\right)$, $\tilde{e}_{L}$,
$\tilde{e}_{T}$,  $\tilde{f}_{T}^{\perp}$, $\tilde{f}_{L}^{\perp}$, $\tilde{h}$
$\tilde{e}_{T}^{\perp}$ are T-odd; and $\tilde{f}^{\perp},\tilde{g}_{T}$ (or $\left.\tilde{g}_{T}^{\prime}\right), $
$\tilde{g}_{T}^{\perp}$, $\tilde{g}_{L}^{\perp}$, 
$\tilde{h}_{L}$,
$\tilde{h}_{T}$,
$\tilde{e}$,
$\tilde{h}_{T}^{\perp}$ are T-even. 
%______________________________edit________________________________________%

These TMDs can be projected out by using disparate Dirac matrices. In the right-hand side of Eq. (5), if one takes the real part then they can derive the traces  of the T-even TMDs. On the other hand, if one uses the imaginary part, T-odd TMDs  are obtained. 
Here we deal specifically with $\tilde{e}_{L}$ and $\tilde{e}_{T}$ \cite{lu2012} by considering the real part of Eq. (5)  and using Dirac matrix  $ i \sigma^{\alpha+} \gamma_{5}$ as

%______________________________________________________________________%

%_________________________Cal 4__Lu2012___edit_______________________________%
\begin{equation}
\frac{1}{2 M x} \operatorname{Tr}\left[\tilde{\Phi}_{A \alpha} i \sigma^{\alpha+} \gamma_{5}\right]=S_{L}\left(\tilde{h}_{L}+i \tilde{e}_{L}\right)-\frac{p_{T} \cdot S_{T}}{M}\left(\tilde{h}_{T}+i \tilde{e}_{T}\right),
\end{equation}
where $S_{T}$ and $S_{L}$ are the transverse and longitudinal polarization vector of the nucleon respectively.
%_______________________________________________________________________%

%______________________________Cal 5__Lu2012_______(Edited)__________________________%

While applying the approximation of the lowest order, we neglect every gauge-link in the correlator (Eq. (4)) and choose the model which has been used extensively in the calculation of TMD distributions \cite{luref22, luref32}. Before all else, diquark model shows that the T-odd distributions are non-vanishing. We consider the case in which the diquark is an axial-vector. 
%______________________________Cal 5__Lu2012____edit______________________________________%
Here, we calculate the T-odd TMDs emerging in the DIS process (Drell-Yan process comes with a minus sign). Above expression is proportional to the sum of the terms $\psi^{* r}_{-+}(x,p_{T}) \psi^{r}_{++}(x,p_{T})$, $\psi^{* r}_{-0}(x,p_{T}) \psi^{r}_{+0}(x,p_{T})$, $\psi^{* r}_{--}(x,p_{T}) \psi^{r}_{+-}(x,p_{T})$, $\psi^{* r}_{++}(x,p_{T}) \psi^{r}_{-+}(x,p_{T})$, $\psi^{* r}_{+0}(x,p_{T}) \psi^{r}_{-0}(x,p_{T})$ and $\psi^{* r}_{+-}(x,p_{T}) \psi^{r}_{--}(x,p_{T})$.
We can solve this further and get the value of desired TMDs by using suitable light-front wave functions \cite{luref6, lfwf1, maji2017}.
Also in a specific Feynman rule \cite{lu2012, luref19, luref20}, the field strength tensor has been used in the form $F^{+\alpha}$ : $-i\left(q^{+} g^{\alpha \rho}-q^{\alpha} g^{+\rho}\right)$. While using this model some divergences are found which appear to be emerging for a few T-odd TMDs when the integrations are performed over transverse momentum. In the quark-diquark model these kind of divergences are explicitly found very often \cite{luref21, lu2012}. So, it can be deduced that T-odd twist-3 distributions has this general feature that when on the nucleon-quark-diquark interaction vertex the point-like coupling is applied the divergences appear. To derive the finite results, one can choose a dipole form factor instead of a point-like coupling constant for the nucleon-quark-diquark coupling \cite{luref22}.

%_______________________________________________________________________%

\section{Conclusion}
%____________________lu 2012______edit_____________________________________________%
%_______________________________________________________________________%
We have investigated the prospect to calculate the T-odd interaction-dependent twist-3 quark distributions in the quark diquark model. In the approxmation of lowest order we find that abandoning the gauge-links in the correlator can give results which are not equal to zero for the eight T-odd interaction-dependent quark TMDs of twist-3. Particularly, we find that the projection of T-odd twist-3 correlator with this specific Dirac matrix $i \sigma^{\alpha+} \gamma_{5}$ in the form of TMDs $\tilde{e}_{L}$ and $\tilde{e}_{T}$
while considering the case of a axial-vector diquark, leads to an equation proportional to the the light-front wave functions and the field strength tensor. From this one can get the expression of each TMD individually by specifying the nucleon helicity.
%_______________________________________________________________________%
%_______________________________________________________________________%
\section*{Acknowledgements}
The work of HD is supported by the Science and Engineering Research Board, Government of India, under MATRICS (Ref No. MTR/2019/000003).

%Acknowledgements should follow immediately after the conclusion.

% TODO: include author contributions
%\paragraph{Author contributions}
%This is optional. If desired, contributions should be succinctly described in a single short paragraph, using author initials.

% TODO: include funding information
%\paragraph{Funding information}
%Authors are required to provide funding information, including relevant agencies and grant numbers with linked author's initials. Correctly-provided data will be linked to funders listed in the \href{https://www.crossref.org/services/funder-registry/}{\sf Fundref registry}.

\bibliography{SciPost_Example_BiBTeX_File.bib}

\nolinenumbers

\end{document}